\documentclass[a4paper]{jpconf}
\usepackage{graphicx}

\usepackage{graphicx}
\usepackage[english]{babel}

\usepackage{amssymb}
\usepackage{hhline}
\usepackage{epsfig}
\usepackage{amsmath}
\usepackage{dcolumn}
\usepackage{multirow}
\usepackage{array}
\usepackage{color}

\newcommand{\be}{\begin{equation}}
\newcommand{\ee}[1]{\label{#1} \end{equation}}
\newcommand{\ba}{\begin{eqnarray}}
\newcommand{\ea}[1]{\label{#1} \end{eqnarray}}

\begin{document}
\title{A 'soft+hard' model for Pion, Kaon, and Proton Spectra and $v_2$ measured in PbPb Collisions at $\sqrt s = 2.76 A$TeV}

\author{Gergely G\'abor Barnaf\"oldi$^{\,1}$, K\'aroly \"Urm\"ossy$^{\,1}$, G\'abor B\'ir\'o$^{\,1,2}$}

\address{$^1$ Wigner Research Center for Physics of the HAS,\\ 29--33 Konkoly--Thege Mikl\'os Str. H-1121 Budapest, Hungary}

\address{$^2$ Lor\'and E\"otv\"os University, 1/A P\'azm\'any P\'eter S\'et\'any, H-1117  Budapest, Hungary}

\ead{barnafoldi.gergely@wigner.mta.hu}

\begin{abstract}
Hadron spectra measured in high-energy collisions present distributions which can be derived from the non-extensive statistical and thermodynamical phenomena. Based on earlier theoretical developments, it seems, the methods are very applicable for jets hadronization processes in electron-positron, proton-proton, and even in heavy-ion collisions. 

Here, we present what can was learnt from the recent theoretical and phenomenological developments: transverse momentum spectra and azimuthal anisotropy ($v_2$) of charge averaged pions, kaons and protons stemming from central Pb+Pb collisions at $\sqrt s$ = 2.76 ATeV are described \textit{analytically} in a `soft + hard' model. 

In this model, we propose that hadron yields produced in heavy-ion collisions are simply the sum of yields stemming from jets (hard yields) in addition to the yields originating from the Quark-Gluon Plasma (soft yields). The hadron spectra in both types of yields are approximated by the Tsallis\,--\,Pareto like distribution.
\end{abstract}


\section{Introduction}
\label{sec:intro}

Perturbative quantum-chromodynamics improved parton model (pQCD) calculations describe the spectra of hadrons stemming from hard processes in proton-proton (pp) collisions~\cite{bib:pQCD,bib:BGG_FF}. Keeping track of parton model calculations at leading order, it can be argued that the Tsallis\,--\,Pareto-like (or cut power-law) distribution is a good approximation for the resulting hadron spectra \cite{bib:Wong1,bib:Wong2}. This observation has been tested by many groups~\cite{bib:phenixPP}--\cite{bib:Pars}. 

In central heavy-ion (AA) collisions, however, the power of the hadron spectra changes dramatically around $p_T\approx$ 6 GeV/c, thus, it cannot be fitted by one single cut power-law distribution. As the $p_T \lessapprox$ 6 GeV/c part of hadron spectra cannot be reproduced by pQCD methods, thermal models based on the Tsallis statistics have been developed for this purpose~\cite{bib:Wilk3}--\cite{bib:BiroJako}. Beyond these models, lies the conjecture that the Quark-Gluon Plasma (QGP) created in high-energy AA collisions has non-conventional thermodynamical properties. Among others, a possible scenario is that negative-binomial (NBD) hadron multiplicity distributions are responsible for the emergence of the non-Boltzmannian (Tsallis\,--\,Pareto like) distribution in high energy reactions~\cite{bib:UKppNdep,bib:BiroNflukt,bib:UKppFF,bib:UKeeFF}.

In this paper, we persue the conjecture presented in Refs.~\cite{bib:UKshAA2,bib:UKshAA3,bib:UKshAA} and make out the hadron spectra measured in AA collisions as the sum of yields stemming from the QGP (we refer to as 'soft yields') and yields coming from jets (we refer to as 'hard yields'). In these papers, transverse spectra and azimuthal anisotropy ($v_2$) of charged hadrons obtained from various centrality Pb+Pb collisions at $\sqrt s = 2.76 A$TeV have been described. In this paper, we turn our attention to the spectra and $v_2$ of identified pions, kaons, and protons created from central Pb+Pb collisions at the same LHC energy.

We note that transverse spectra and $v_2$ of various identified hadrons measured at RHIC energy have been described by a similar model~\cite{bib:Tang1,bib:Tang2}. In that model, spectra measured in pp collisions have been used as hard yields, and it has been conjectured that hard yields are suppressed at low $p_T$.

\section{Fits to transverse spectra and $v_2$}
\label{sec:fitspec}

We made fits on recent data measured in Pb+Pb collisions at $\sqrt s = 2.76 A$TeV by the ALICE experiment. We used the identified hadron spectra from Ref.~\cite{bib:ALICEdNdpT} as a basis of our 'soft+hard' fits and then these parameters were converted to obtain asymuthal anisotropy data. Our $v_2$ results were compared to the measured values based on Ref.~\cite{bib:ALICEv2}. 

\begin{figure}[h]
\begin{center}
\includegraphics[width=0.9\textwidth]{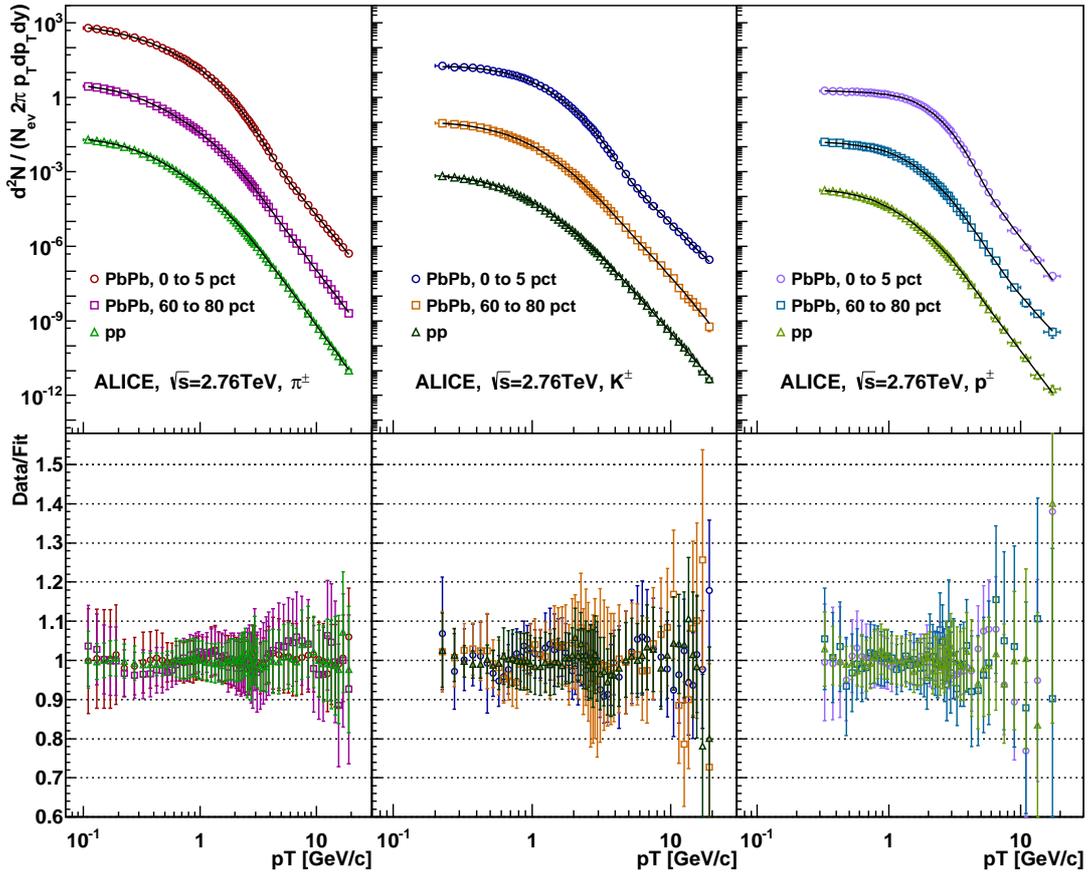}
\end{center}
\caption{{\sl Top panels}: Transverse momentum spectra of charge averaged pions, kaons and protons stemming from central ($0-5\%$) and periferal ($60 \%-80\% $) Pb+Pb as well as pp collisions at $\sqrt s = 2.76 A$TeV~\cite{bib:ALICEdNdpT}. Curves are fits of Eq.~(\ref{eq12}). {\sl Bottom panels:} Data over fit plots.\label{fig:dNdpT}}
\end{figure}

\newpage
We conjecture that hadrons stemming both 'soft' and 'hard' yields are distributed according to differently parametrized Tsallis distribution: 
\be
\left. \frac{dN}{2\pi p_T dp_T dy} \right|_{y=0} = \sum\limits_i f_i = \sum_i A_i \left[1 + \frac{(q_i-1)}{T_i}[\gamma_i(m_T-\nu_{0,i} p_T) - m] \right]^{-1/(q_i-1)}\;,
\ee{eq12}
with $i$ = 'soft' or 'hard', $m_T = \sqrt{p_T^2 + m^2}$, $\nu_{0,soft}$ being the transverse flow velocity of the QGP on the freeze-out hyper-surface as given in Ref.~\cite{bib:UKinAuAu}, and $\nu_{0,hard}$ being an average jet velocity (see e.g. in Ref.~\cite{bib:UKppFF}), finally, $\gamma_i = 1/\sqrt{1-\nu_{0,i}^2}$. These yields have maxima at $p^{max}_{T,\,i} = \gamma_i\, m\, \nu_{0,i}$. Due to the  pion mass, these maxima are below the measurement range in the case of pions and charged averaged hadrons, and thus $\nu_{0,i}$ cannot be determined accurately. The argument in Eq.~(\ref{eq12}) may be approximated by $[\gamma_i(m_T - \nu_{0,i} p_T) -m]/T_i \approx p_T/T^{Dopp}_i $ with the Doppler-shifted parameters $ T^{Dopp}_i = T_i\,\sqrt{\left(1+\nu_{0,i} \right)/\left(1-\nu_{0,i}\right)} $, which be compared to experimantal data.

Fits were plotted on Fig.~\ref{fig:dNdpT}: transverse momentum spectra of charge averaged pions, kaons and protons stemming from central ($0-5\% $) and peripheral ($60 \%-80\% $) Pb+Pb as well as pp collisions at $\sqrt s = 2.76 A$TeV~\cite{bib:ALICEdNdpT}. Curves are fits of Eq.~(\ref{eq12}). Data/fit ratios were also presented at the lower row, which agreement supports the validity of the theory within $10\%$. 
\begin{figure}[h]
\begin{center}
\includegraphics[width=0.4\textwidth]{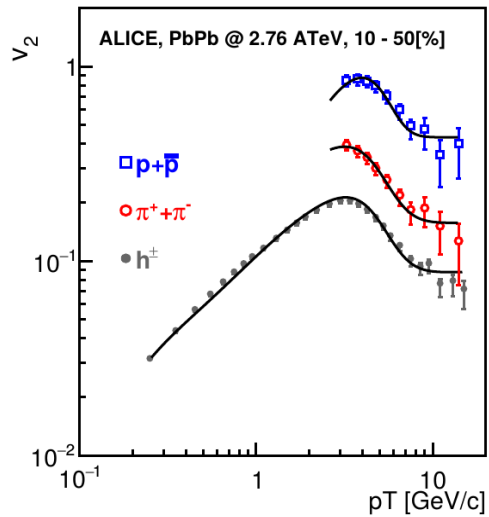}
\end{center}
\caption{The $v_2$ of charge averaged pions, protons and charged hadrons stemming from $10\%-50\%$ centrality Pb+Pb collisions at $\sqrt s$ = 2.76 ATeV \cite{bib:ALICEv2}. Curves are fits of Eq.~(\ref{eq17}).\label{fig:v2}}
\end{figure}
As discussed in Refs.~\cite{bib:UKshAA3,bib:UKshAA}, Tsallis-like hadron distributions and small modular fluctuations of the transverse flow velocities as a function of the azimuthal angle $\nu_i(\varphi) = \nu_{0,i} + \sum_{n=1}^{\infty} \delta \nu_{n,i} \cos(n\varphi)$, result in an azimuthal anisotropy of the measured one for $10-50\%$ central Pb+Pb on Fig.~2:
\be
v_2 = \frac{w_{hard} \,f_{hard} + w_{soft}\, f_{soft}}{f_{hard} + f_{soft}}\; ,
\ee{eq17}
where the coefficient functions are
\be
w_i = \frac{\delta \nu_{2,i}\, \gamma^3_i}{2T_i} \frac{p_T - \nu_{0,i}\, m_T}{1 + \dfrac{q_i-1}{T_i}\,\big[\gamma_i(m_T - \nu_{0,i}\,p_T) - m \big]} \;.
\ee{eq18}
Fits of Eqs.~(\ref{eq12})~and~(\ref{eq17}) are shown in Figs.~\ref{fig:dNdpT} and~\ref{fig:v2} with their parameters are shown in Tab.~1. These are in agreement, what was obtained for the 'soft' component: $\nu_{0,soft} = 0.693 \pm 0.051$ and  $ \delta \nu_{2,soft} = 0.053 \pm 0.017$, as well as, we got for charge averaged hadrons in Refs.~\cite{bib:UKshAA3,bib:UKshAA}. 
\begin{table}[h]
\begin{center}
\begin{tabular}{llccccc}
\hline 
Type & Hadron & $q_i$ & $T_i$ [GeV] &  $A_i$ & $\nu_{0,i}$  \vspace{2mm} \\
\hline \hline
 & $\pi^{\pm}$ & $1.046 \pm 0.016$ & $0.260 \pm 0.025$ & $1252. \pm 324.$ & --- \\
Soft 
 & $p/\bar{p}$ & $1.006 \pm 0.057$ & $0.153 \pm 0.024$ & $4.453 \pm 0.83$ & $0.71 \pm 0.0337$ \\
\hline
 & $\pi^{\pm}$ & $1.172 \pm 0.010$ & $0.091 \pm 0.012$ & $2947. \pm513.$ & --- \\
Hard 
 & $p/\bar{p}$ & $1.133 \pm 0.031$ & $0.238 \pm 0.0712$ & $9.18 \pm 1.82$ & --- \\
\hline 
\hline
\end{tabular}
\caption{Parameters for the charge averaged indentified pion and proton spectra.}
\end{center}
\label{tab:par}
\end{table}

\section{Conclusions}
\label{sec:concl}

The 'soft+hard' model were presented and applied on ALICE data in pp and Pb+Pb collisons.

\ack
\label{sec:ack}

This work was supported by Hungarian OTKA grants K104260, NK106119, and
NIH TET 12 CN-1-2012-0016. Author GGB also thanks the J\'anos Bolyai
Research Scholarship of the H.A.S.


\end{document}